\documentclass[onecolumn,amsmath,amssymb]{revtex4}

\usepackage[english]{babel}
\usepackage[utf8x]{inputenc}
\usepackage{amsmath}
\usepackage{graphicx}
\usepackage[colorinlistoftodos]{todonotes}

\begin{document}

\title{Analytical Modelling of the Spread of Disease in Confined and Crowded Spaces}

\author{Lara Gosc\'e}
\author{David A.W.\ Barton}
\author{Anders Johansson}
\email{a.johansson@bristol.ac.uk}
\affiliation{Faculty of Engineering, University of Bristol, UK}

\date{\today}

\begin{abstract}

Since 1927, until recently, models describing the spread of disease have mostly been of the SIR-compartmental type, based on the assumption that populations are homogeneous and well-mixed. The focus of these models have typically been on large-scale analysis of scenarios such as cities, nations or even world scale. SIR models are appealing because of their simplicity, but their parameters, especially the transmission rate, are complex and depend on a number of factors, which makes it hard to predict how a change of a single environmental, demographic, or epidemiological factor will affect the population. Therefore, in this contribution we start to unpick the transmission-rate parameter. Analysing the implications that arise when taking crowd behaviour explicitly into account, we show how both the rate of infection as well as the walking speed depend on the local crowd density around an infected individual. The combined effect is that the rate of infection at a population scale has an analytically tractable non-linear dependency on crowd density. We model the spread of a hypothetical disease in a corridor and compare our new model with a typical compartmental model, which highlights the regime in which current models may not produce credible results.

\end{abstract}

\maketitle

\section*{INTRODUCTION}

During its history, humanity has repeatedly encountered major epidemics such as the Plague (Black Death) in the 14\textsuperscript{th} century, the Spanish Flu in 1918 or more recently the Swine Flu (2009), to mention only a few. 
Some of these epidemics severely affect the global population, such as the Plague that in only six years caused the death of almost one third of the European population.

One of the first analytical approaches to study and model the systemic nature of the spread of an infectious disease, was Kermack and McKendrick \cite{kermackmckendrick} who proposed what would later been modified and defined as an SIR model, that divided a population into three compartments; (i) susceptible, $S$; (ii) infectious, $I$; and (iii) recovered/removed, $R$, and devised a set of differential equations to model the transition rates from each of these states to the others. This seminal work would lead to the onset of a new field of Mathematical Epidemiology. By introducing also an (iv) exposed class, $E$, the model becomes what is called an SEIR model \cite{li}:
\begin{equation}
\begin{array}{rl}
\dot S  =& - \lambda IS / N\\
\dot E =& \lambda IS/N - k E\\
\dot I =& k E - \alpha I\\
\dot R =& \alpha I
\end{array}
\label{eq_sir_unnormalised}
\end{equation}
where $S+E+I+R=N$ (population size).

To make this model more practical for different population sizes, we will now normalise the model according to $s=S/N$, $e=E/N$, $i=I/N$, and $r=R/N$ which results in Eq.~\eqref{eq_sir}:
\begin{equation}
\begin{array}{rl}
\dot s =& - \lambda is\\
\dot e =& \lambda is - k e\\
\dot i =& k e - \alpha i\\
\dot r =& \alpha i
\end{array}
\label{eq_sir}
\end{equation}
where $\lambda$ is the {\em per-capita} transmission rate, $\alpha$ is the constant recovery rate (thus, the mean infectious period is $1/\alpha$) and $1/k$ is the mean exposed period.

The SEIR model is based on the following four main assumptions:
\begin{enumerate}
\item The population size $N$ is constant.
\item There is no heterogeneity, i.e. the impact of individual characteristics such as age, sex and behaviour are neglected.
\item The presence of an exposed class means that the susceptible individuals are not instantly able to infect while being infected themselves, but they will be after a fixed amount of time.
\item The population is well-mixed. This means that each individual has the same probability of contracting the disease.
\end{enumerate}

Therefore, such an approach will neglect the importance of heterogeneity in the social interaction and mobility patterns of individuals. The validity of the `well-mixed' assumption is particularly problematic. Recent Statistical Physics approaches however, have incorporated such elements into disease-spreading models \cite{brockmann,vespignani} which has resulted in higher fidelity in predictions of disease outbreaks.

A growing body of literature on detailed studies of crowd behaviour in crowded places have revealed a number of insights into non-linear, dynamic, adaptive and self-organised crowd behaviour that seriously question the validity of the well-mixed assumption in the original SIR model, even if network effects are taken into account to achieve some degree of heterogeneity.

Since the beginning of the $21^{st}$ century, mathematical epidemiology has found in the use of complex networks a new and more precise way to tackle the problem; where individuals are represented by nodes and their interactions by links. Moreover the nodes have a certain connectivity distribution i.e. the probability $P(k)$ that a node is connected to $k$ other nodes. This connectivity is typically either exponential (in case of exponential networks) or of power-law (in case of scale-free networks). An important result was obtained in Ref.~\cite{pastor} where it as shown that exponential networks show an epidemic threshold that separates an infected phase from a healthy one, while the behaviour of scale-free networks depends on the power-law exponent and may or may not represent the epidemic threshold.

Extensive research has been focused on the study of epidemics through network theory. However, while contemporary network science approaches \cite{netwbr, netwnew, moreno} to modelling the spread of disease has greatly improved the modelling accuracy at large scales, there is still limited understanding of the effect of crowd behaviour that come into play in densely crowded and confined spaces such as transport hubs, mass gatherings, and city centres. 

In recent years the study of human interaction and mobility has provided some important insights that are needed to increase the realism of epidemic models, mostly at large scales \cite{dirk}. In particular, sensors such as mobile phone tracking, WiFi/Bluetooth devices, and CCTV are used to acquire important information of how people move in their daily lives as well as during mass gatherings or other crowded events. 

On an even smaller scale, interesting experiments are being carried on in small and crowded environments (conference halls, hospital wards, museums etc) by using wearable sensors (such as RFID devices) that are capable of sensing face-to-face interaction of individuals \cite{cattuto, stehle, barrat, van}. These studies have shown the importance of heterogeneity and dynamic contact patterns in shaping the dynamics of the infection, by the use of simulations, temporal networks and contact matrices. However, we still lack an analytical instrument that allows us to find a common design and evaluation strategy of the problem in general cases.

Therefore, in this contribution, we propose a way in which insights into crowd behaviour can be used to improve prevailing compartmental models (well known for their simplicity). A microscopic foundation is particularly important to evaluate potential epidemiological implications of designs or operational plans for one-off events where we do not have the luxury of existing data that can be analysed and used as a basis for an epidemiological model. For recurring events, our study will provide an important analytical toolkit that can be used to study the epidemiological implications of changing one of the crowd-related design or operational parameters, and keeping everything else constant.

By analysing a tractable scenario -- people moving in a corridor -- and building an SEI model, we focus our attention at one of the most elusive parameters of mathematical epidemiology: the contact rate. We will study how contact rate depends on crowd behaviour, and most importantly, the crowd density.

\section*{RESULTS}

We recognise that human mobility in crowded places exhibit a rich set of complex, adaptive and self-organised behaviours~\cite{transci}. Therefore, we propose a bottom-up modelling approach to the spread of disease in crowded places, starting with a simple but analytically tractable special case, that can later be extended to take a richer set of crowd behaviours and more complex environments into account.

\subsection*{Traditional SEIR model in a corridor}

\begin{figure}
\begin{center}
\includegraphics[width=0.7\textwidth]{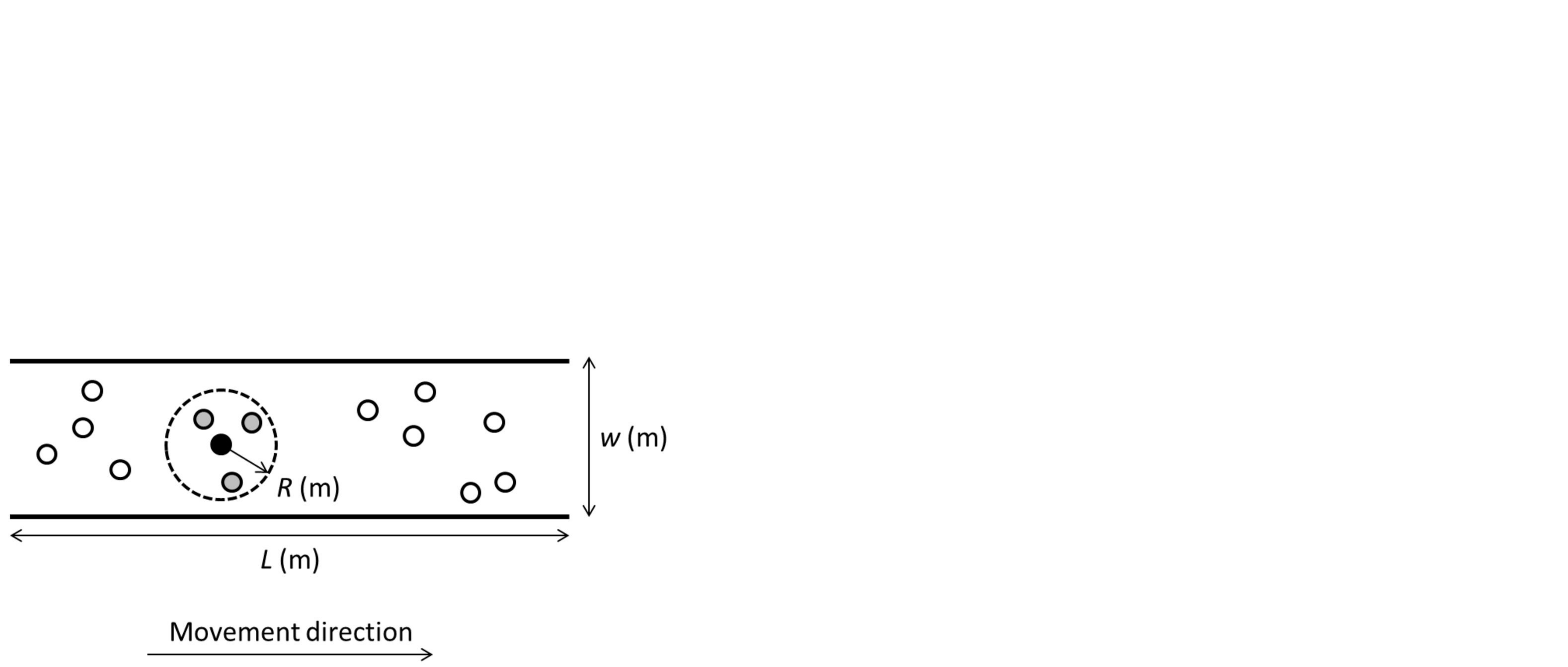}
\caption{Illustration of corridor with periodic boundary conditions, length $L$ metres and width $w$ metres. The small black circle shows a single infected individual, the larger dashed circle shows the radius $R$ of infection, the smaller grey circles show individuals within the radius of possible infection, and the small white circles show individuals that are currently outside the reach of possible infection.}
\label{fig_corridor}
\end{center}
\end{figure}

Before we introduce our new methodology taking crowd behaviour into account, let us start by implementing the traditional SEIR model in a corridor with unidirectional crowd flow. The length of the corridor is $L$ metres, the width is $w$ metres and the corridor has periodic boundary conditions (see Fig. \ref{fig_corridor}). In this environment, $1\%$ of the population is infected at time $0$. Adding to the previous SEIR model assumptions, we also have the following corridor-specific assumptions:
\begin{enumerate}
\setcounter{enumi}{4}

\item Periodic boundary conditions (and thus, a constant population size $N$) is assumed in the corridor.

\item The time scale (i.e. maximum simulated time) is smaller than the recovery time and exposed time (i.e. only the exposed and susceptible classes will be affected within simulated time). This also means that we will not take secondary infections into account.



\end{enumerate}

Now, since we aim to {\em improve} rather than to replace current models, we will put our model into prevailing terminology, and thus, each of the individuals in our model can take any of the states (i) susceptible, $s$; (ii) infectious, $i$; and (iii) exposed, $e$, which represents individuals that have become infected after contact with an infected person but who are not yet infectious due to the relatively short duration of our simulation, for this same reason the compartment $r$ is not present. We will thus obtain the normalised SEI model described by Eq.~\eqref{eq_sie}.
\begin{equation} 
\begin{array}{rl}
{\dot s}  =& - \lambda i s \\
{\dot e} =& \lambda i s \\
{\dot i} =& 0 \\
\end{array}
\label{eq_sie}
\end{equation}
where $i_0 = 0.01$.

\subsection*{Improved density-dependent SEIR model}

Since we want to take crowd behaviour explicitly into account, we will now propose an improved model where $\lambda$ will depend on the spatio-temporal distribution of the crowd density. 

Earlier work \cite{xie, zhu, zhao} has studied distances that large droplets are carried; (i) 6 m away by exhaled air at a velocity of 50 m/s (sneezing); (ii) 2 m away at a velocity of 10 m/s (coughing); and (iii) 1 m away at a velocity of 1 m/s (breathing).

We will build on these findings and therefore replace the earlier assumption 4 (i.e. {\em 'The population is well mixed'}) with the following: Infections can only happen within an $R$-metre radius of an infected individual. The final new assumption that the new model will be based on is:
\begin{enumerate}
\setcounter{enumi}{6}
\item The mixing rate (i.e. the arrival rate of new pedestrians) around infected individuals due to variable walking speeds is much higher than the rate of infection, which means that we do not need to take local saturation effects into account.
\end{enumerate}

In previous work \cite{nichol,li}, the per-capita transmission rate $\lambda$ has been defined as a constant, reflecting the {\em average} number of contacts within a population. Such an approach can lead to accurate outbreak predictions only if $\lambda$ is calibrated to specific cases, but it does not incorporate the effect of crowd behaviour which can lead to characteristically different contact rates in different scenarios. We also assume that the mixing rate (i.e. the rate of arrival of new pedestrians) around infected individuals due to variable walking speeds is much higher than the rate of infection (which means that we do not need to take local saturation effects into account).

In our case, each infected individual can only reach $n$ other individuals within a 1-metre radius (see Fig. \ref{fig_corridor}).

In reality, this number $n(t)$ is time dependent, $n(t)=\rho(t) \pi R^2$, since it depends on the local density $\rho(t)$ (1/m${^2}$) of the crowd surrounding each infected individual, multiplied by the area $\pi R^2$ where an infection can take place.

Taking time-dependent crowd behaviour into account would make our model (Eq.~\eqref{eq_sie}) too complicated for practical purposes. Therefore, we will re-formulate crowd-dynamic theory into a statistical description that we can later incorporate into our existing model (Eq.~\eqref{eq_sie}). To do so, we will start with a previously discovered Eulerian description \cite{headway} of the crowd density distribution $p^s$, and re-formulate this into a Lagrangian description $p^t$ that will better capture the effect that pedestrians spend longer time in crowded areas due to their lower walking speed in such areas.

We know from Ref.~\cite{headway} that if a random location is picked somewhere within an open space with area $A$ (m$^2$) and the {\em average} density $\varrho$ (of the whole space) is $\varrho = N/A$ (m$^{-2}$), the local density $\rho$ (m$^{-2}$) can be described by a Gamma distribution, according to Eq.~\eqref{eq_gamma}.
\begin{equation}
p^{s}(\rho;A;B) = \frac{B^A}{\Gamma(A)}\rho^{A-1} e^{-B\rho}
\label{eq_gamma}
\end{equation}
where $A=3 \mu = 3 \varrho = 3N/(Lw)$, $\varrho$ is the global density of the whole corridor, and $B=3$.

Moreover, if a random pedestrian $i$ (in our case an infected individual) is picked at a random time $t$, the local density $\rho$ around that individual $i$ will follow the distribution described by Eqs.~\eqref{eq_pt} and \eqref{eq_cases}.
\begin{equation}
p^{t}(\rho; A; B) = \frac{p^{s}(\rho; A; B)}{\langle v(\rho)\rangle}
\label{eq_pt}
\end{equation}
where: 
\begin{equation}
0.05 \le \langle v(\rho)\rangle = \frac{1/\sqrt{\rho} - 1/\sqrt{\rho_\text{max}}}{0.5} \le 1.34
\label{eq_cases}
\end{equation}
is the average velocity (1.34 m/s is the average value of the free speed when the corridor is empty \cite{headway}) and $\rho_\text{max}$ is the maximum value of the density.

\begin{figure}
\begin{center}
\includegraphics[width=0.65\textwidth]{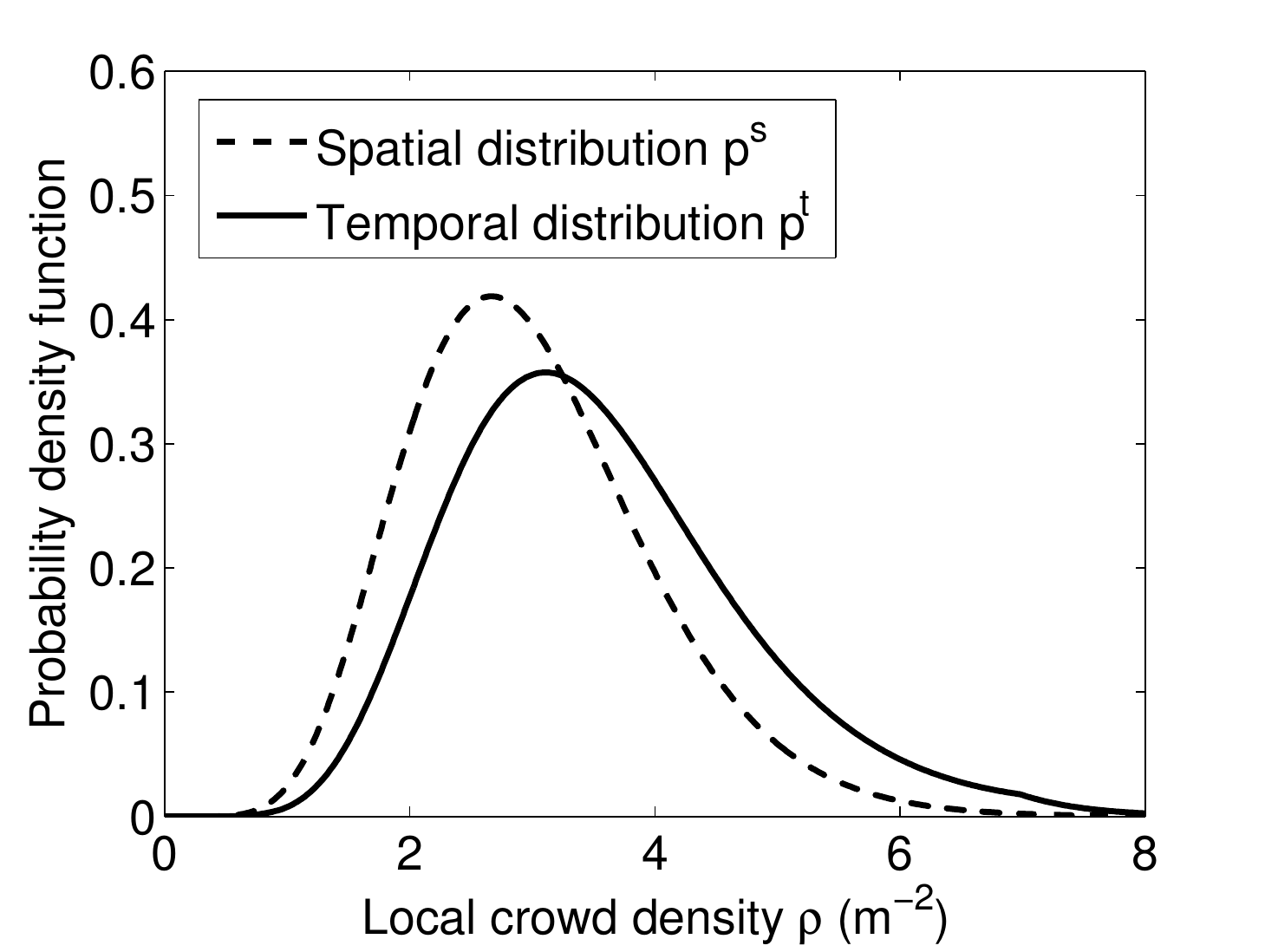}
\end{center}
\caption{Probability-density of the spatial distribution $p^s$ of the local crowd density $\rho$ in a space with an average crowd density $\varrho=N/A = 3$ m$^{-2}$ as an example, compared with the (spatio-)temporal distribution $p^t$ of local density across the same space. The difference is due to a lower walking speed in high density areas, which will lead to individuals spending longer time in dense areas compared to lower density areas.}
\label{fig_distributions}
\end{figure}

The probability density functions of the crowd density over space ($p^s$) and space-and-time ($p^t$) are illustrated in Fig.~\ref{fig_distributions}. The reason why $p^t$ is skewed towards larger crowd density values compared to $p^s$ is because pedestrians walk slower when they are in dense regions compared to less dense regions, which means that they will spend more time in crowded areas compared to less crowded areas.

Because walking speed is a decreasing function of crowd density, individuals spend relatively longer periods of time in crowded areas compared to less crowded areas. Therefore, from an individual perspective, the average value of the local density around an infected individual is given by Eq.~\eqref{eq_inddens}.

\begin{equation}
\langle \rho \rangle = \int_0^{\infty} \ p^{t}(\rho)\rho\, d\rho
\label{eq_inddens}
\end{equation}
which means that, substituting this value in the definition of the contact rate we obtain:
\begin{equation}
\lambda^\text{density-dependent} = c \times \underbrace{\pi R^2 \times \int_0^{\infty} \ p^{t}(\rho)\rho\, d\rho}_\textrm{Number of pedestrians in circle}
\label{eq_lambda_dens_dep}
\end{equation}
where $c$ is the transmissibility i.e. the fraction of infected-susceptible contacts that actually lead to an infection.

In this way, we have obtained a new description of the transmission rate between pedestrians in a crowded location, a value that depends on the crowd density. With the previous model definition, the value of the transmission rate did not depend on the crowd density in the corridor which is the case for the new model.

\subsection*{Numerical results}

\begin{figure}
\includegraphics[width=0.7\textwidth]{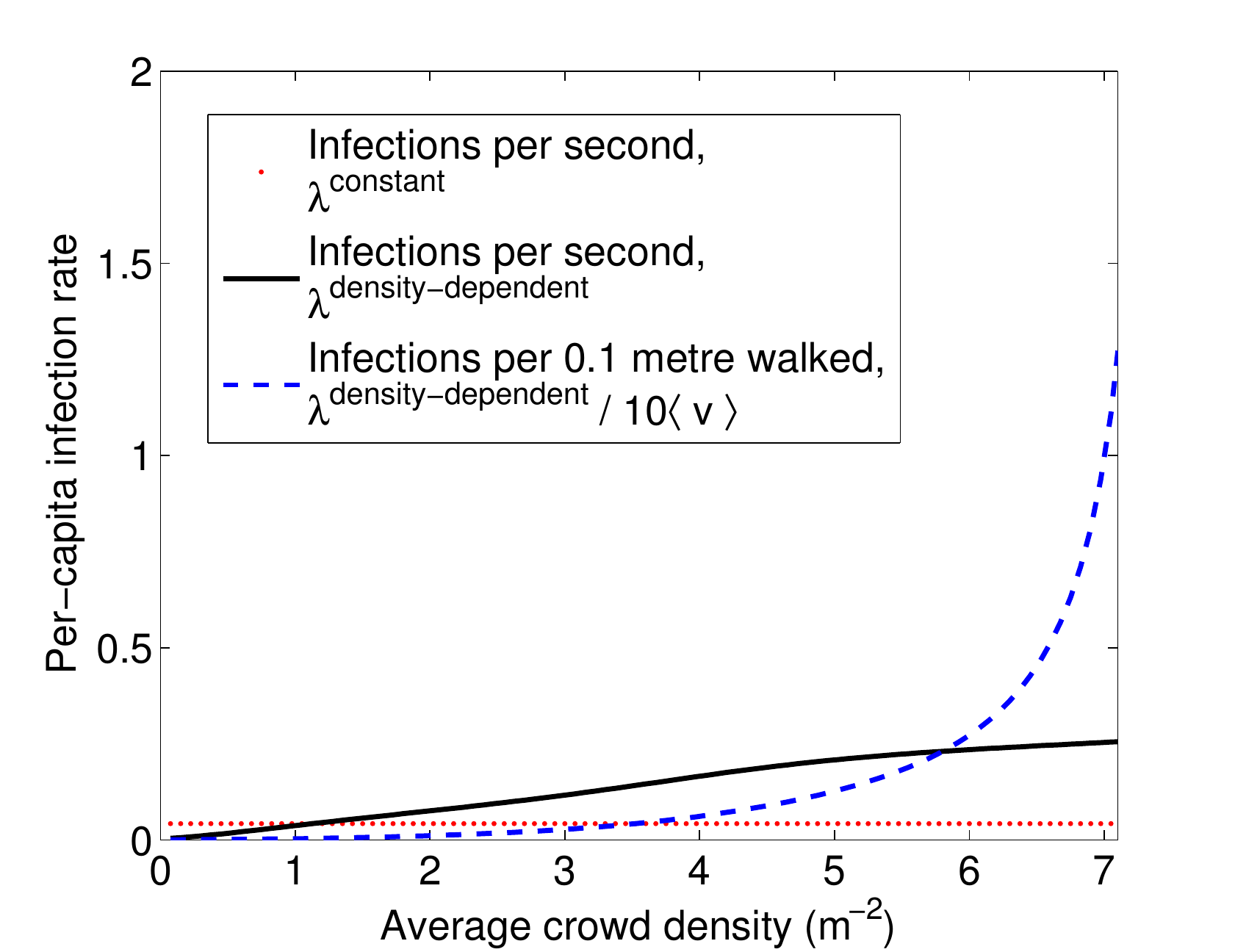}
\caption{Dependency of per-captita infection rate $\lambda$ on average crowd density $\varrho=N/A$. Note that the rate of infection per 0.1 metre walked, peaks at high crowd density due to the decreasing walking speed at high crowd density.}
\label{fig_varrho_vs_lambda}
\end{figure}

Let us now compare our new density-dependent model with the previous model to investigate what effect a crowd-density-dependent per-capita infection rate $\lambda^{density-dependent}$ will have on the population scale. As can be seen in Fig. \ref{fig_varrho_vs_lambda}, the rate of infection per unit time is a non-linear function of crowd density. For practical purposes, the infection rate {\em per distance walked} is a better metric than the infection rate {\em per unit time}. For example in a busy transport hub, people have to walk a fixed distance from one platform to another regardless of the level of crowdedness and will therefore spend longer time altogether in that transport hub on crowded days (thus, being even more affected by prevalent diseases in that space).


To show the population-scale impact of our proposed model, we simulate a hypothetical disease that spreads between two individuals when they are at close proximity. We set the constant per-capita transmission rate to $\lambda^\text{constant}=0.01$. If we take $\langle \rho \rangle = 1.0$ m$^{-2}$ as an arbitrary crossover point between $\lambda^\text{constant}$ and $\lambda^\text{density-dependent}$, we get $c = 0.01 / (\pi R^2)$ which will define $\lambda^\text{density-dependent}$ according to Eq.~\eqref{eq_lambda_dens_dep}.

\begin{figure}
\begin{center}
\includegraphics[width=0.7\textwidth]{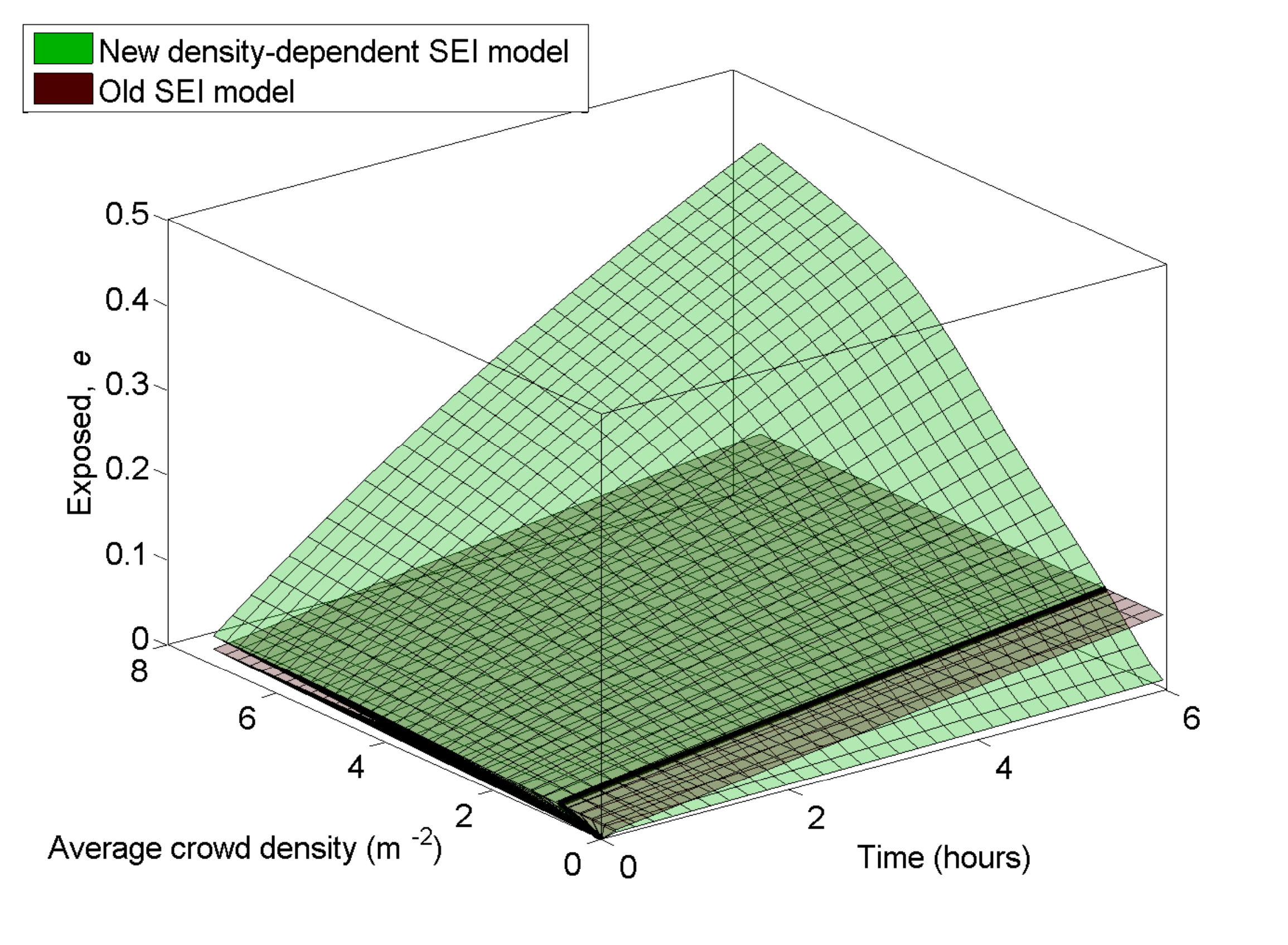}
\end{center}
\caption{Numerical results showing how the fraction of individuals in the exposed state $e$ depends on time $t$ and average crowd density $\varrho=N/A$. The two surfaces correspond to a traditional compartmental model (referred to as `old SEI model' and the model we propose in this paper referred to as `density-dependent SEI model').}
\label{fig_surfaces}
\end{figure}

We now run our old and new density-dependent SEI models next to each other for different values of average density in the interval $\varrho = N/A \in [0, 8]$ (m$^{-2}$). We then plot the model results for the fraction of exposed individuals in the population, for a population size $N=1000$, as a function of both time $t$, and average crowd density $\varrho$ (see Fig.~\ref{fig_surfaces} for results). It is clear that even though the old and the newly proposed model give comparable results within a narrow range of crowd densities, for large portions of the density spectrum, the old model either severely under- or over-estimates the rate of infection of the disease.

\pagebreak

\subsection*{Model validation}

\begin{figure}
\begin{center}
\includegraphics[width=0.7\textwidth]{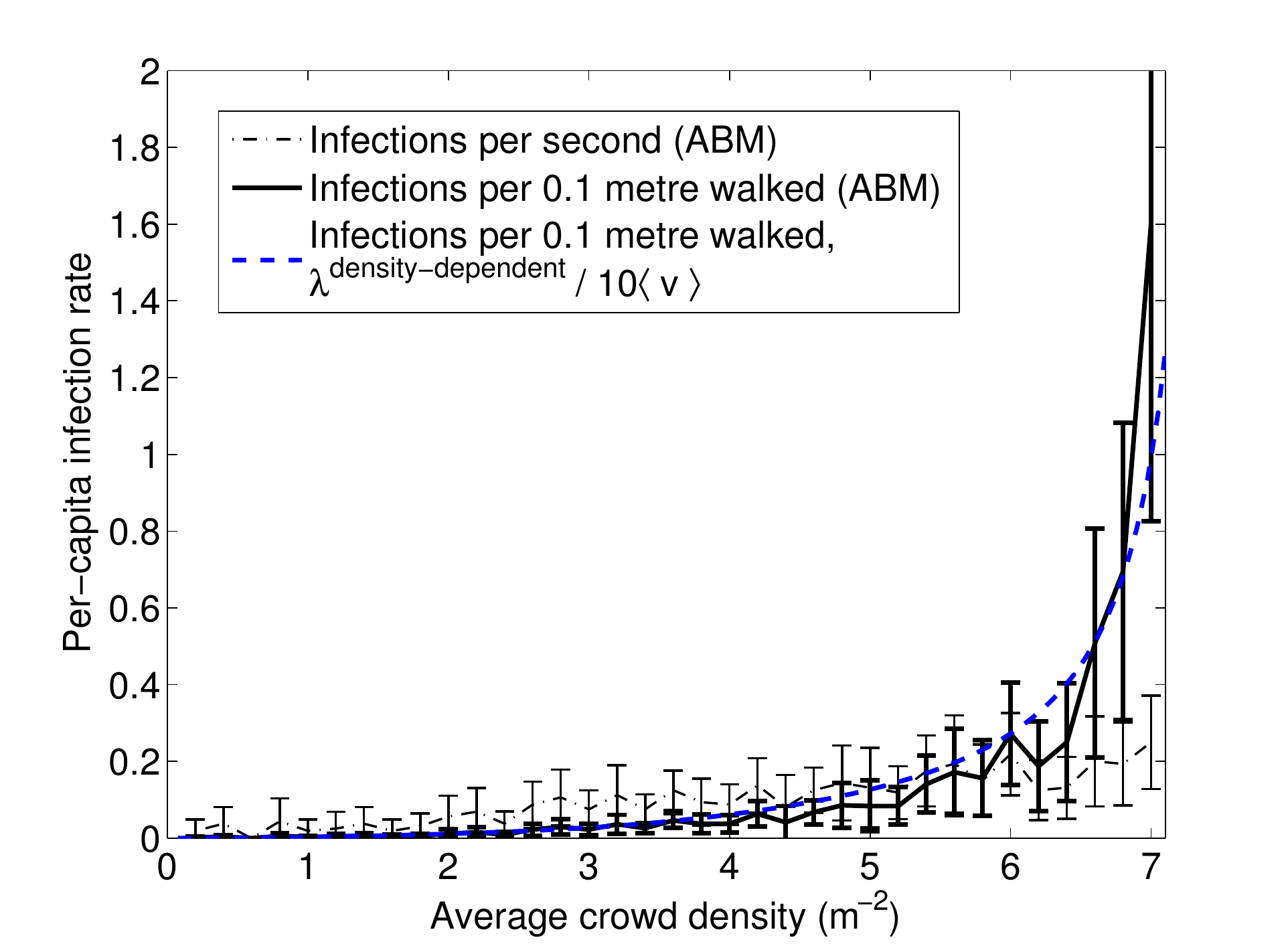}
\includegraphics[width=0.7\textwidth]{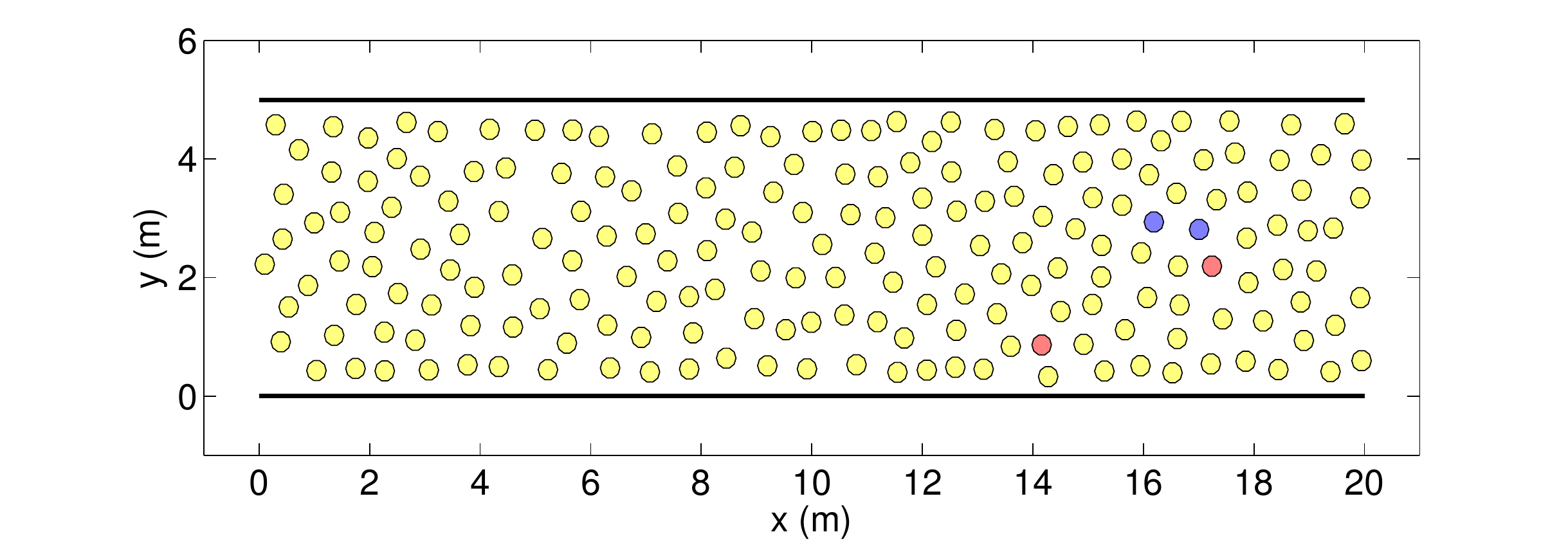}
\caption{Top: Results of the agent-based model of a unidirectional crowd flow in a corridor. The curves correspond to the mean values of 10 simulations for a number of average crowd densities, and the error bars correspond to 1 standard deviation. The blue dashed curve is the same as in Fig.~\ref{fig_varrho_vs_lambda} for comparison. Bottom: Snapshot of the agent-based crowd model. Colours correspond to: susceptible (yellow); infected (red); and exposed (blue).}
\label{fig_abm}
\end{center}
\end{figure}

A comprehensive quantitative model validation is not within the scope of this paper. However, to be able to validate the main characteristics of our model, we will run an agent-based crowd model of unidirectional crowd flow in a corridor. Our crowd model is based on the {\em social-force model} \cite{wenjian} where the motion of pedestrian $i$ is described by
\begin{equation}
	m_i \frac{d \vec{v}_i(t)}{dt} = \vec{f}_i(t)
	\label{eq_abm_first}
\end{equation}
where $m$ is the mass of the pedestrian (in kg), $\vec{v}$ is the walking velocity, and $\vec{f}_i$ is a force described by
\begin{equation}
	\vec{f}_i(t) = m_i \frac{1}{\tau} (v_i^0 \vec{e}_i - \vec{v}_i(t) ) + \sum_{j \ne i} \vec{f}_{ij}(t) + \sum_{k} \vec{f}_{ik}(t)
\end{equation}
where $\vec{f}_{ij}$ is a repulsive force from pedestrian $j$ acting on pedestrian $i$ specified as

\begin{equation}
	\vec{f}_{ij} = F exp \big( -d_{ij}/D_0 \big) \vec{e}_{ij}
	\label{eq_fij}
\end{equation}
where $d_{ij}$ is the distance from pedestrian $i$ to pedestrian $j$, $\vec{e}_{ij}$ is the normalised vector pointing from the centre of mass of pedestrian $j$ to the centre of mass of pedestrian $i$.

The force $\vec{f}_{ik}$ from boundary $k$ onto pedestrian $i$ is also specified by Eq.~\eqref{eq_fij}, but the position of pedestrian $j$ is replaced by the closest point on boundary $k$.

The model parameters used are the same~\footnote{Note that our model does not use the second force component $(D_1/d_ij)^k$ used to accurately model turbulent and very dense crowds, since that is not the focus of our paper.} as in Ref.~\cite{wenjian}: The free speeds $v^0$ are Gaussian distributed with mean 1.34 m/s and standard deviation 0.26 m/s when pedestrians are unobstructed. However, speeds are also bounded by $v_i^0 < d^{min}_i(t) / T_i$ where $d^{min}_i(t)$ is the minimum net-distance to all surrounding pedestrians $j$ and $T_i$ is the preferred net-time headway that are Gaussian distributed with mean 0.5 s and standard deviation 0.1 s.

The mass $m$ of pedestrians are Gaussian distributed with mean 60 kg and standard deviation 10 kg; the maximum pair-wise repulsive force is set to $F$=160 N; the relaxation time is set to $\tau$=0.5 s; and the remaining parameter $D_0$ that determines the typical length scale of the repulsive forces is set to $D_0=0.31$ m.


The results of the agent-based model (see Fig.~\ref{fig_abm}) matches the main characteristics of our model results as shown in Fig.~\ref{fig_varrho_vs_lambda}. A more comprehensive model validation will be carried out in future work.

\section*{DISCUSSION AND FUTURE WORK}

Our study shows a way in which detailed insights gained from data-driven crowd research can be utilised to improve current disease spreading models in an analytically tractable way. Thus, the spread of disease depends strongly on the behaviour of the crowd, and in particular the contact rate is a key parameter in the study of the evolution of a disease and it varies considerably depending on the density of the population in the studied environment.

We have found three different but related effects of crowd behaviour on the rate of infection in a confined space. All three effects contribute to an increased infection rate for increasing crowd density; (i) increasing crowd density leads to decreased proximity which leads to a higher number of individuals within the range of infection around an infected individual; (ii) density-dependent walking speeds lead to individuals spending longer periods of their time in crowded locations compared to less crowded locations within an area with locally varying crowd densities [thus, the {\em mean} crowd density underestimates the typical proximities between individuals]; and (iii) for the same reason, individuals also spend longer time {\em altogether} in environments such as transport hubs on crowded days compared to less crowded days, which leads to an increasing rate of infection {\em per distance walked}.

\medskip

Comparing our new model with existing models, we have shown how existing models may give significant over- or under-estimations of the spread of disease in certain settings, and we have also shown how current models can easily be improved by a simple modification of the $\lambda$ parameter.

Moreover, the use of a generic definition of the transmission rate that depends on the transmissibility of the disease, the area considered and the density of the population, is a powerful tool that can be used in many different crowd scenarios, and will decrease the dependency on a-posteriori empirical transmission rate data every time one of the environmental or crowd-related parameters have changed.

Having a more precise prediction of what may, or may not, happen in crowded environments in terms of contagion between individuals could be highly useful to be able to devise adaptive control strategies during mass gatherings, in busy transport hubs, or in busy city centres.

\section*{METHODS}

Figure~\ref{fig_distributions} was produced by numerically solving Eqs.~\eqref{eq_gamma} -- \eqref{eq_cases} for $\rho$ on the interval [0.01, 8] with a step size of 0.01.

Note that the average walking speed $\langle v(\rho) \rangle$ is truncated at an upper boundary at 1.34 m/s (i.e. the 'free' unobstructed walking speed), and by a lower boundary at 0.01 (m/s). All intermediate walking speeds are calculated using the non-linear function specified by Eq.~\eqref{eq_cases}.

Figure~\ref{fig_varrho_vs_lambda} was produced by solving Eq.~\eqref{eq_lambda_dens_dep} for 100 values of $\rho$ equally distributed on the interval [0.01, 8].

Figure~\ref{fig_surfaces} was created by solving Eq.~\eqref{eq_sie} for a population size $N=1000$, using our proposed density-dependent specification of $\lambda$ obtained from Eq.~\eqref{eq_lambda_dens_dep} and comparing the results to the number of exposed individuals produced using a constant value $\lambda^{constant}$. The range of crowd density is exactly the same as the one used in Fig.~\ref{fig_varrho_vs_lambda}. The time axis uses a time step of 60 seconds and the time range used is [0, 6 $\times$ 3600] seconds (i.e. six hours).

Figure~\ref{fig_abm} was obtained by repeatedly running an agent-based crowd model 10 times for each of 40 different values of crowd density in the range [0, 8] m$^{-2}$ using $N$=200 pedestrians in a corridor with periodic boundary conditions. The model ran for 10 simulated seconds and data from $t < 2$ s was excluded because of the time needed for pedestrians to accelerate from their initial zero speeds at the start of the simulation. The model equations~\eqref{eq_abm_first}~--~\eqref{eq_fij} were solved numerically using the 1st order Euler method with a time step of 0.05 s.


\section*{ADDITIONAL INFORMATION}

\subsection*{Author Contributions}

AJ, DB, and LG designed research; LG and AJ performed research and wrote the manuscript text; AJ, DB and LG reviewed and approved the manuscript.

\subsection*{Competing Financial Interests}

The authors declare no competing financial interests.

\end{document}